# Drawing WS$_2$ thermal sensors on paper substrates

Martin Lee[a], Ali Mazaheri[b,c], Herre S. J. van der Zant[a], Riccardo Frisenda[b], Andres Castellanos-Gomez[b,*]

[a.] *Kavli Institute of Nanoscience, Delft University of Technology, Lorentzweg 1, 2628 CJ Delft, The Netherlands..*
[b.] *Materials Science Factory. Instituto de Ciencia de Materiales de Madrid (ICMM-CSIC), Madrid, E-28049, Spain.*
[c.] *Nanophysics research Lab., Department of Physics. University of Tehran, Tehran 14395, Iran.*

\* Andres.castellanos@csic.es

Paper based thermoresistive sensors are fabricated by rubbing WS$_2$ powder against a piece of standard copier paper, like the way a pencil is used to write on paper. The abrasion between the layered material and the rough paper surface erodes the material, breaking the weak van der Waals interlayer bonds, yielding a film of interconnected platelets. The resistance of WS$_2$ presents a strong temperature dependence, as expected for a semiconductor material in which charge transport is due to thermally activated carriers. This strong temperature dependence makes the paper supported WS$_2$ devices extremely sensitive to small changes in temperature. This exquisite thermal sensitivity, and their fast response times to sudden temperature changes, is exploited thereby demonstrating the usability of a WS$_2$-on-paper thermal sensor in a respiration monitoring device.

## Introduction

The research field of paper-based-electronics is advancing tremendously fast, spurred by the societal, industrial and technological demands of ultra-low-cost electronic components.[1,2] In fact, paper substrates are several orders of magnitude cheaper than convention silicon wafers enabling for a considerable reduction of the component costs.[1] Moreover, paper substrates are biodegradable, which is an important feature for its applicability in the emerging field of disposable electronic components, [1,3–5] and hypoallergenic, a crucial property for its use as substrate in electronic components for healthcare applications. These two properties of paper substrates, in combination with its ultra-low-cost, makes paper supported devices very interesting to substitute reusable sensors like thermometers, pulse oximeters or breathing monitoring devices (known vectors for infectious outbreaks[6–8]) in healthcare applications.

The downside of paper-electronics, however, is the fact that the rough (fibrous) surface of paper hampers the use of traditional device fabrication protocols which are developed and optimized for silicon based electronic components. This handicap can be overcome by exploiting the roughness of paper surface to erode materials while rubbing them against the paper substrate.[3,9,10] This method, similar to writing or drawing with a pencil on paper, allows for the deposition of continuous films of material on the paper surface. Up to now this method have been extensively used to fabricate several sensors and electronic devices based on graphite [9,11–18] and very recently it has been extended to other van der Waals materials like semiconducting transition metal dichalcogenides.[19,20]

Here, we exploit the strong temperature dependence of the resistance of semiconducting WS$_2$ to fabricate thermal sensors on paper substrates. We find a strong drop of the device resistance upon a temperature increase, as expected from thermally activated charge carrier transport in a semiconductor material, which makes these devices extremely sensitive to small changes in temperature. We further demonstrate the fast response to sudden temperature changes, and we illustrate the potential of these $WS_2$-on-paper thermal sensors in respiration monitoring applications.

## Device fabrication and characterization

The device fabrication is a slight variation with respect to that recently reported in Ref. [20] for the fabrication of $MoS_2$-on-paper photodetectors. Briefly, the device outline is printed with toner with an office laser printer (Brother MFCL5700DN) on standard 80 gr/m$^2$ copier paper (Fig. 1a and 1f). The channel area is delimited by masking around with Scotch tape (Magic tape 3M, see Fig 1b). A cotton swab is then used to rub $WS_2$ micronized powder (HAGEN automation Ltd) against the bare channel area (Fig. 1c and inset in 1g). After rubbing the $WS_2$ powder, the channel tape mask is removed (Fig. 1d and 1g) and the electrodes are then drawn on top of the $WS_2$ channel with a 4B pencil (~80% graphite content [21], see Fig. 1e). Given the dimensions of a sharp pencil tip it is rather straightforward to draw straight lines with a width smaller than 300 µm. Moreover, following the printed electrodes outline one can reliably obtain interdigitated electrodes with ~ 600 µm gap. Figures 1f-1h show pictures of a $WS_2$ device with interdigitated graphite electrodes on paper at different fabrication stages. The inset in Fig. 1g shows a picture of a cotton swab used to rub the $WS_2$ powder.

We use scanning electron microscopy (SEM) and energy dispersive X-ray spectroscopy (EDX), using a FEI Helios G4 CX system, to characterize the morphology and the composition of the $WS_2$ powder and the resulting $WS_2$ films after rubbing it on paper. Copper tape was used to electrically ground the samples during the measurement. An electron energy of 30 keV was used for imaging and EDX spectroscopy.

Figures 2a–b show SEM images of the as-received micronized $WS_2$ powder where one can see that it is formed by platelets with typical lateral dimensions in the ~2-3 µm range and ~100-200 nm in thickness. Fig. 2c–d are SEM images of the drawn $WS_2$ film on paper showing that $WS_2$ is deposited on top of the paper fibers and filling in the gaps between them. Note that, due to the insulating character of paper, uncoated paper would have a huge difference in contrast in the SEM images. As during the SEM characterization, we did not observe spots with a large contrast difference we conclude that the $WS_2$ rubbing yields an even deposition on the surface of the paper. However, there are differences in the morphology between the $WS_2$ deposited onto the paper fibers and the one deposited filling in the gaps between fibers. On top of the fibers the resulting film seems to be a compact composite of small crystallites with only few loose platelets on the surface. We recently observed a similar behaviour for $MoS_2$ films deposited on paper by rubbing $MoS_2$ single crystals and we attributed it to the larger abrasion forces and pressure on the surface of the paper fibers that crushes the material in small crystallites pressed together to form a compact film.[20] The gaps between the paper fibers, on the other hand, are filled with $WS_2$ flakes of slightly smaller lateral dimensions than those present in the pristine powder (Fig. 2d). This is expected as the flakes would be subjected to lower friction forces and pressures within these cavities between fibers during the rubbing process.

We have also characterized the chemical composition with EDX spectroscopy (Fig. 2e). Both the $WS_2$ powder and the $WS_2$ film on paper presents two prominent peaks at ~1.8 keV and ~2.3 keV that corresponds to the W and S elements, respectively. The $WS_2$ on paper spectrum also shows an enhancement of the peak corresponding to O and the appearance of a C peak, due to the signal of the underlying paper.

We used EDX mapping to get an insight into the interface between $WS_2$ film on paper and the graphite electrode drawn on top of the $WS_2$. A 4B pencil is used to draw a graphite film onto a $WS_2$-on-paper film. Then graphite/$WS_2$/paper stack is sliced with a sharp razor blade to obtain a clean cut that allows to study the cross-sectional structure of the stack. The sliced sample is finally mounted on the SEM sample holder with the freshly sliced edge laying in the image plane (with < 5° of uncertainty). Figure 3a shows the SEM image of the cross-section where the vacuum, the film on paper, the paper substrate and the SEM sample holder can be resolved. In Fig. 3b the same SEM image is shown but now false-coloured (using the results of the EDX maps discussed below) to highlight the different components. Figure 3c displays the EDX map corresponding to the O element, which arises due to the presence of the paper substrate. The corresponding EDX map of the C element in Fig. 3d indicates two different regions: one area that matches with the presence of the paper substrate and a higher intensity region on the top of the film. In fact, on the topmost part of the film/paper cross-section there is a ~5 µm thick layer with brighter C signal that we attribute to the presence of the topmost graphite layer. An intensity line cut in the C element map clearly illustrates the presence of this ~5 µm thick layer with brighter C signal. The EDX maps of the W and S elements are shown in Figs. 3e-f and the presence of the $WS_2$ film with an approximate thickness of ~20-30 µm is clearly visible. Interestingly, by overlapping the bright C intensity area of Fig. 3d (attributed to the graphite topmost layer) and the bright W and S areas of Fig. 3e-f one can see how the $WS_2$ and graphite films are well-separated (see Fig. 3b). We address the reader to a complementary Raman study in the Supporting Information studying the cross-section composition of the drawn heterostructure films on paper. This capability of creating well-defined heterostructures between different van der Waals materials by simply 'drawing' materials on top of each other has been also demonstrated by Withers and co-workers.[19]

## Results

**Thermoresistive characteristics of the $WS_2$-on-paper devices**

We found that $WS_2$ devices with the geometry described in Fig. 1 typically have resistances in the 0.7 MΩ to 20 MΩ range and thus can be easily read-out with a conventional handheld multimeter (Bolyfa 117 USB). In fact, one can reach conductivities up to $(3.5 \pm 1.3) \cdot 10^{-3}$ S/m (corresponding to an electrical resistivity of $285 \pm 77$ Ω·m), that is ~30-40 times higher conductivity than that of films prepared by spray coating liquid phase exfoliated $WS_2$ on PET.[22] The conductivity is also 20 times higher than that found for $MoS_2$ on paper [20] and 100 times larger than the higher conductivity values reported for conductive networks of liquid phase exfoliated $MoS_2$.[23–25] Nonetheless the conductivity of the $WS_2$-on-paper films is still lower than the in-plane conductivity of bulk (~$10^2$ S/m)[26] and mechanically exfoliated single-layer $WS_2$ (~$10^3$ S/m),[27] indicating a large contribution of the out-of-plane transport as well as the flake-to-flake hopping in the total conductivity of the $WS_2$ films on paper. In the Supporting Information a transfer length measurement on a $WS_2$ film on paper is described to extract the contact resistance and the conductivity. Note that we attribute the device-to-device dispersion in resistance to the strong dependence of the resistance on the film thickness and the density of percolative conduction paths.

To benchmark the temperature sensitivity of the $WS_2$-on-paper thermal sensors we have measured the electrical resistance as a function of the temperature. Figure 4 shows the temperature dependence of the resistance of a $WS_2$-on-paper device. The resistance drops abruptly upon temperature increase. the downward trend cannot be fitted to a single exponential decay, indicating the presence of several thermally activated processes. The inset in Fig. 4 plots the relative change in resistance (with respect to the resistance at room temperature, $T = 25°C$) of the $WS_2$ device and compares it with that measured on a graphite device (rubbing Pressol®, finely ground natural graphite flakes, on the channel instead of $WS_2$ powder), which has been previously proposed as a prospective material for thermoresistive sensors on paper [10].

The change in resistance for the WS$_2$ device is much stronger than that observed in the graphite device, indicating that WS$_2$-on-paper thermal sensors present a higher thermal sensitivity.

In order to quantitatively compare the temperature sensitivity of different thermoresistive sensor one can use the temperature coefficient of resistance (TCR) as a figure of merit, which is defined as:

TCR = (d$R$/d$T$)/$R$ .

For the WS$_2$ devices the TCR ranges from -20 000 ppm/°C to -160 000 ppm/°C while the graphite device has a TCR ranging from -2 500 ppm/°C to -3 700 ppm/°C, the latter being in good agreement with previously reported values.[13,28] Therefore, the WS$_2$ devices show a remarkably higher temperature sensitivity than materials previously used in paper-supported thermal sensors like graphite and carbon nanotube yarn (-700 ppm/°C),[29] and even higher than commonly used temperature sensing metals such as platinum (3 920 ppm/°C), copper (4 300 ppm/°C) and nickel (6 810 ppm/°C).[30] Table 1 summarizes the TCR values determined here for van der Waals materials and those reported in the literature.

In order to further characterize the response speed and reproducibility of our WS$_2$-on-paper thermal sensors we warm up a WS$_2$ device to $T \sim 45$ °C and suddenly change the temperature by blowing cool air (at $T$ = 20-22 °C and with ~30-40% or relative humidity) on its surface with a hand-pump to inflate balloons. Figure 5a shows the time evolution of the device resistance highlighting the moment when the hand pump blew cool air onto the device. The resistance increases suddenly upon blowing, at a response time comparable to the instrumentation acquisition time ~ 0.2 s, indicating indeed a decrease of temperature. After the sudden jump, the resistance decays exponentially until thermalizing again with the thermal bath. The recovery time of the sensor can be estimated as ~ 1 s, limited by the specific heat and thermal conductivity of the film + paper system. Figure 5b shows the time evolution of the device resistance while subjecting the device to 21 blow/thermalization cycles to illustrate the reproducibility of the device. The right axes in Fig. 5a and 5b displays the actual temperature of the device, obtained from a resistance *vs*. temperature curve like the one shown in Fig. 4, showing that upon cool air blowing the device temperature drops 8-10 °C. Note that in this temperature range, and blowing air with relatively low humidity, we can rule out any moisture condensation/evaporation at the surface of the device during this experiment. In fact, under these conditions the calculated dew point is between 5 and 10 °C, much lower than the device surface temperature. We address the reader to the Supporting Information for a similar test carried out on a graphite device to illustrate the superior thermal sensitivity of WS$_2$ with respect to graphite for thermal sensing on paper substrates.

**WS$_2$-on-paper thermal devices for respiration monitorization**

We illustrate one potential application of these paper-based thermal sensors as respiration monitoring devices to detect the inhaling and exhaling steps of the breathing cycle through the associated changes in temperature. Figure 6 show the time evolution of the resistance of a WS$_2$ device while a test-user breathes at ~ 1 cm distance over it. As we found that the WS$_2$ thermoresistive sensors are also very sensitive to humidity changes and the response is very slow (~5 s), we encapsulated the WS$_2$ thermoresistive device with Scotch tape to exclude the effect of the humidity changes induced during the breathing cycles and thus to obtain reproducible readouts. When the test-user starts breathing over the sensor the resistance drops by ~8%, corresponding to a temperature increase of ~1.5 °C. During the breathing monitoring the resistance oscillates because of the temperature fluctuations induced by the inhaling/exhaling cycles. The magnitude of the resistance change reaches ~2-3%, which is notably larger than that observed in graphite-based (~0.5%)[29] and carbon nanotube-based (~0.2-0.3%)[31] thermoresistive breathing sensors. Moreover, these WS$_2$-on-paper thermal sensors can be a good alternative to capacitive based breathing monitoring sensors that require more sophisticated electronics for the read-out.[32,33]

## Conclusions

In summary, we introduce paper-supported thermal sensors based on $WS_2$. We demonstrate that the resistance of $WS_2$-on-paper displays an exquisite thermal sensitivity, much higher than that of graphite and other commonly used materials in thermoresistive applications, as expected for a semiconductor material with thermally activated transport. We show that the devices respond fast to sudden changes in temperature (~ 0.2 s) with recovery times of ~ 1 s. Finally, we illustrate the potential of these paper-based devices in breathing monitoring applications. The processes shown in this work could be easily expanded to other (semiconducting) van der Waals materials opening up new possibilities of paper-supported sensors based on other layered materials.

## Acknowledgements

This project has received funding from the European Research Council (ERC) under the European Union's Horizon 2020 research and innovation programme (grant agreement n° 755655, ERC-StG 2017 project 2D-TOPSENSE). R.F. acknowledges the support from the Spanish Ministry of Economy, Industry and Competitiveness through a Juan de la Cierva-formación fellowship 2017 FJCI-2017-32919. A.C-G., M.L. and H.v.d.Z acknowledge the support from the European Union's Horizon 2020 research and innovation program under the Graphene Flagship (grant agreement number 785219, GrapheneCore2 project and grant agreement number 881603, GrapheneCore3 project). We acknowledge support of the publication fee by the CSIC Open Access Publication Support Initiative through its Unit of Information Resources for Research (URICI).

## Notes and references

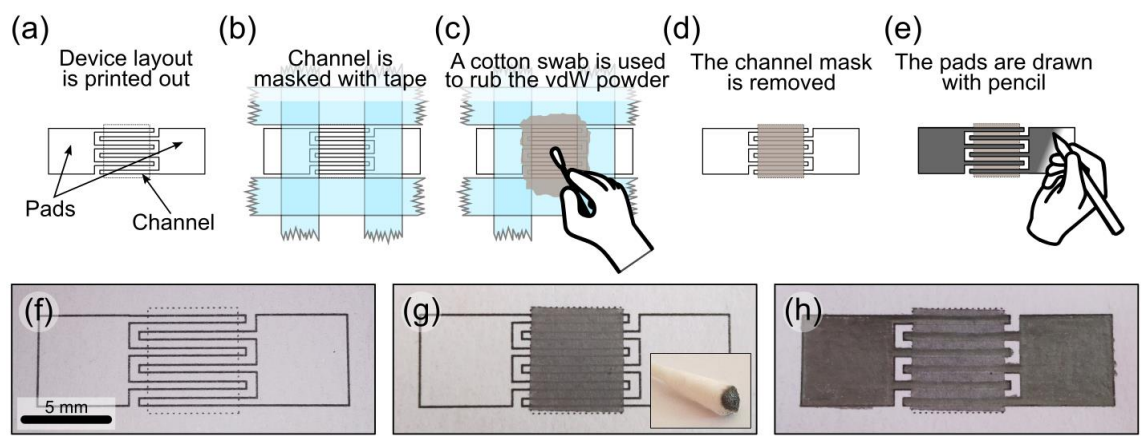

**Figure 1.** Fabrication of thermal sensors with van der Waals materials on paper substrates. (a) The device layout is printed on standard copier paper with a laser printer. (b) The channel area is delimited with Scotch tape mask. (c) A cotton swab is used to rub the powder of van der Waals material against the bare paper area. (d) The mask is removed. (e) The electrodes are drawn with a pencil. (f) Picture of the device layout printed on copier paper. (g) Picture of a deposited $WS_2$ film on the channel area. (Inset) Picture of a cotton swab used to rub $WS_2$ micronized powder. (h) Picture of a finished $WS_2$ device with interdigitated graphite electrodes.

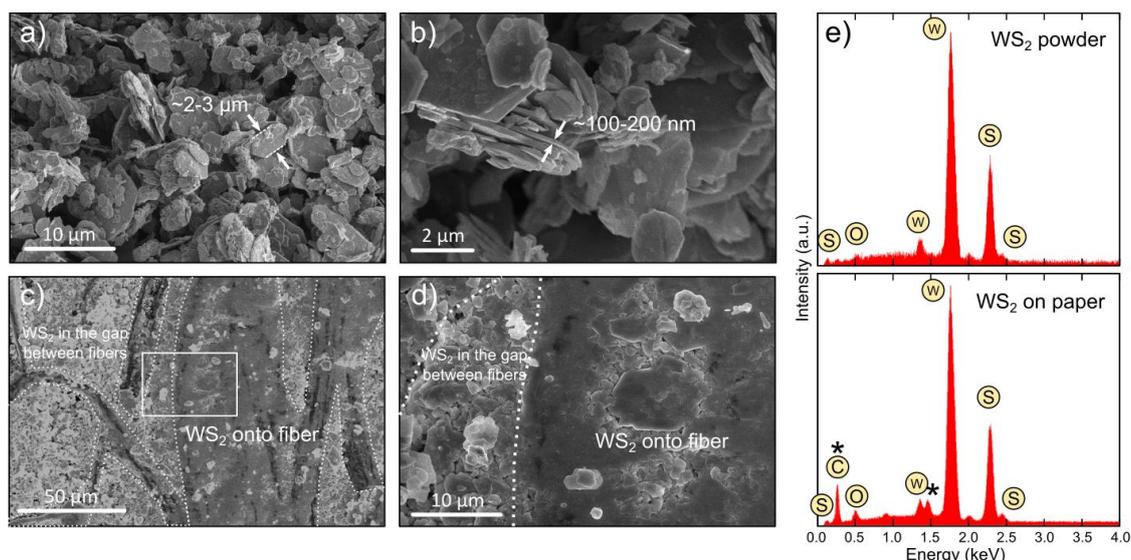

**Figure 2.** Scanning electron microscopy (SEM) and energy dispersive X-ray (EDX) spectroscopy analysis of the micronized $WS_2$ powder and of a $WS_2$ film drawn on paper. (a-b) SEM images of the micronized $WS_2$ powder showing the morphology of the platelets: the typical dimensions of the flakes are ~2-3 µm lateral size and a thickness of ~100-200 nm. (c) SEM image of a $WS_2$ film drawn on copier paper where different domains are resolved: $WS_2$ deposited into the gap between paper fibers and $WS_2$ deposited onto the paper fibers. (d) Zoomed-in image of the region highlighted with a white rectangle in (c) to illustrate the difference in morphology of the $WS_2$ deposited onto the fibers and into the gaps between fibers. (e) EDX analysis of the chemical composition of the $WS_2$ powder (top) and film on paper (bottom). The features highlighted with * correspond to the underlying paper substrate.

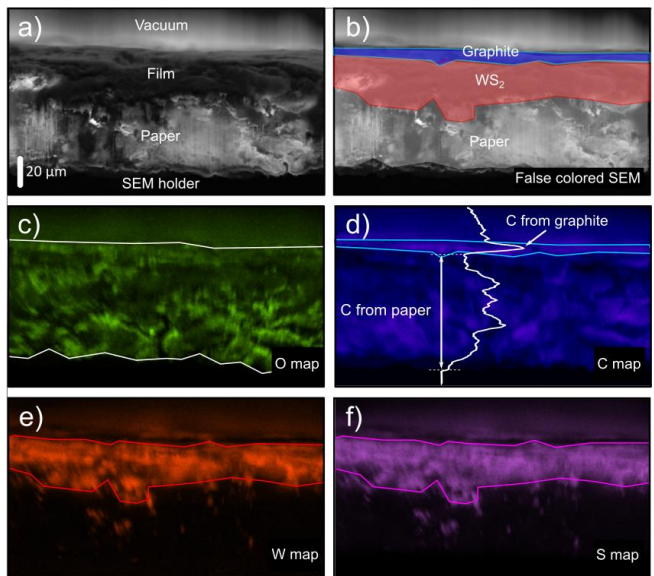

**Figure 3.** Cross-section scanning electron microscopy (SEM) and energy dispersive X-ray (EDX) spectroscopy of a graphite electrode drawn onto the WS$_2$ film on paper. (a) SEM image of the copier paper cross-section with a WS$_2$/graphite heterostructure film drawn onto its surface. (b) False colored SEM image highlighting the different materials composing the heterostructure (extracted from the EDX maps). (c-f) EDX maps for oxygen, carbon, tungsten and sulphur elements, respectively. In the carbon map a vertical intensity line profile has been included to highlight the presence of a higher intensity carbon signal arising from the topmost graphite film, drawn on top of the WS$_2$ film.

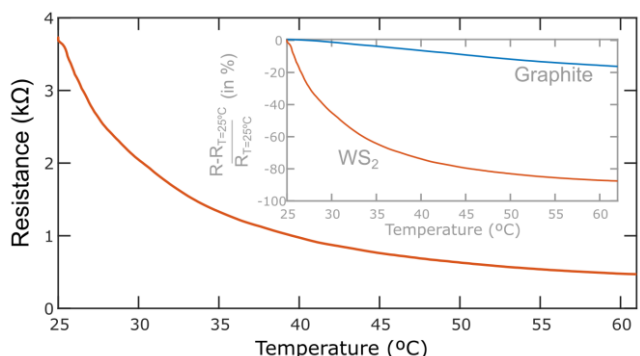

**Figure 4.** Temperature dependence of the resistance of WS$_2$-on-paper thermoresistive devices. (a) Resistance vs. temperature for a WS$_2$-on-paper device with pencil-drawn interdigitated devices. (inset) Comparison between the resistance vs. temperature curves measured for the WS$_2$ and a graphite thermoresistive devices on paper.

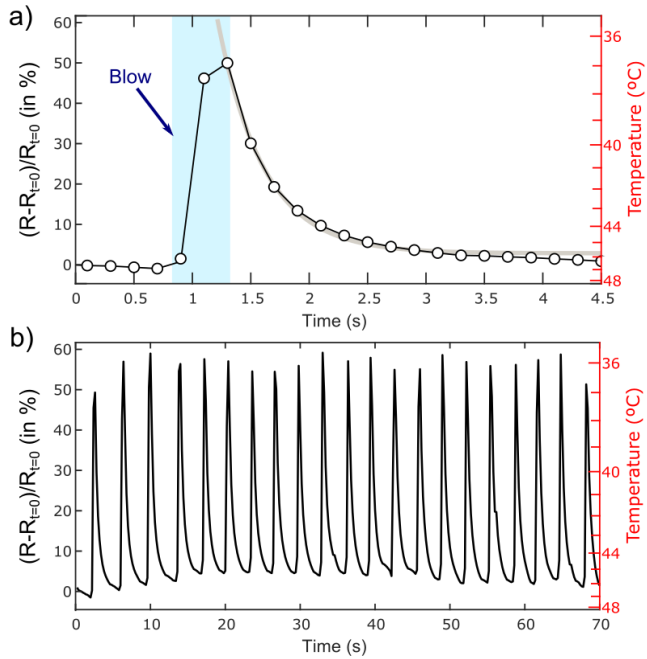

**Figure 5.** Response time of WS$_2$-on-paper thermoresistive devices upon sudden temperature changes. (a) Resistance change vs. time of a WS$_2$-on-paper device subjected to a sudden change in temperature. The device is warmed up to ~45 °C and at time t ~ 1 s a hand pump blows cool air onto the device. (b) 21 blowing cycles to demonstrate the reproducibility of the thermoresistive device.

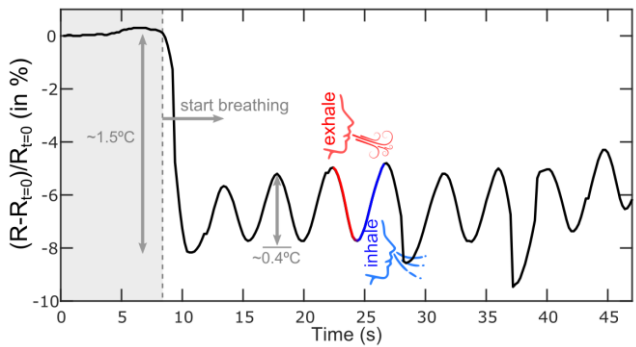

**Figure 6.** Implementation of a respiration monitoring device using the WS$_2$-on-paper thermal sensor. Resistance change vs. time of WS$_2$ thermal sensor device when a test person breaths at ~1 cm over the sensor (breathing starts at t ~ 8 s). The thermoresistive device resolves the inhaling/exhaling cycles because of a ~0.3-0.5 °C increase in the temperature of the sensor upon exhaling. Note: the sensor has been encapsulated with Scotch tape to reduce the effect of humidity changes during the breathing cycles on the resistance.

**Table 1.** Comparison between the temperature coefficient of resistance (TCR) measured here for van der Waals based thermal sensors on paper and other materials reported in the literature.

| Material | TCR (ppm/°C) | Reference |
|---|---|---|
| WS$_2$ | -20 000 to -160 000 | This work |
| Graphite | -2 500 to -3 700 | |
| Graphite | -2 900 to -4 400 | [13] |
| Graphite | -3600 | [28] |
| CNT yarn | -700 | [29] |
| Platinum | 3 920 | |
| Copper | 4 300 | [30] |
| Nickel | 6 810 | |

Supporting Information

# Drawing WS$_2$ thermal sensors on paper substrates

Martin Lee,[1] Ali Mazaheri[2,3], Herre S. J. van der Zant[1], Riccardo Frisenda[2], Andres Castellanos-Gomez[2,*]

[1]Kavli Institute of Nanoscience, Delft University of Technology, Lorentzweg 1, 2628 CJ Delft, The Netherlands.

[2]Materials Science Factory. Instituto de Ciencia de Materiales de Madrid (ICMM-CSIC), Madrid, E-28049, Spain.

[3]Nanophysics research Lab., Department of Physics. University of Tehran, Tehran 14395, Iran.

andres.castellanos@csic.es

**Comparison between the performance of WS$_2$ and graphite on-paper thermoresistive devices**

**Measurements on different WS$_2$ devices**

**Optical characterization of the WS$_2$ film on paper**

**Raman characterization of the graphite electrode drawn on top of the films**

**Characterization of the electrical properties of a WS$_2$ film on paper**

**Resistance vs. temperature of device shown in Figure 4 in log scale**

# Comparison between the performance of WS$_2$ and graphite on-paper thermoresistive devices

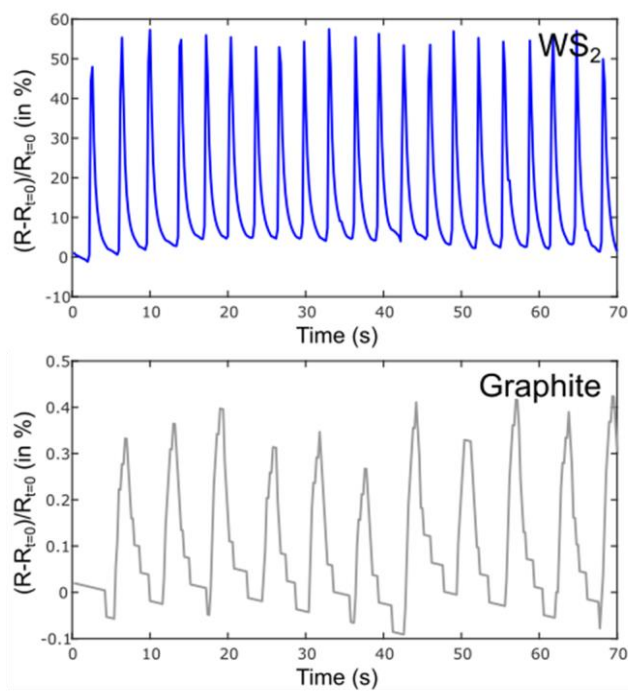

**Figure S1. Comparison between the response of WS$_2$ and graphite on-paper thermoresistive devices upon sudden temperature changes.** The devices are warmed up at ~45-55 °C and a hand pump blows pulses of cool air onto the device that show up as increases in resistance. Note the big difference in the vertical axis for the different materials.

# Measurements on different WS$_2$ devices

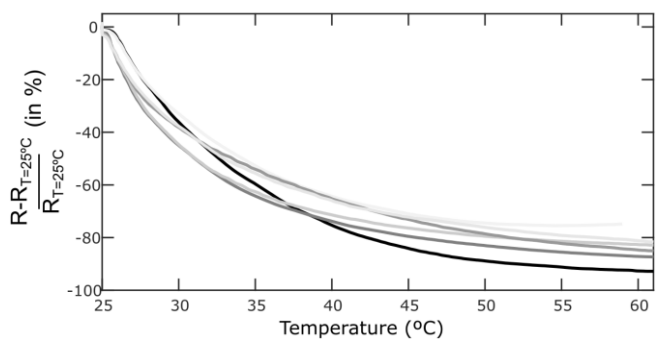

**Figure S2. Change in the resistance vs. temperature for 6 different WS$_2$ devices.** Note that the devices were intended to span over a wide range of thicknesses. In fact, their room temperature resistance varies between 0.7 MΩ and 20 MΩ. Despite the difference in device resistance the resistance change is rather reproducible in all the devices.

# Optical characterization of the WS₂ film on paper

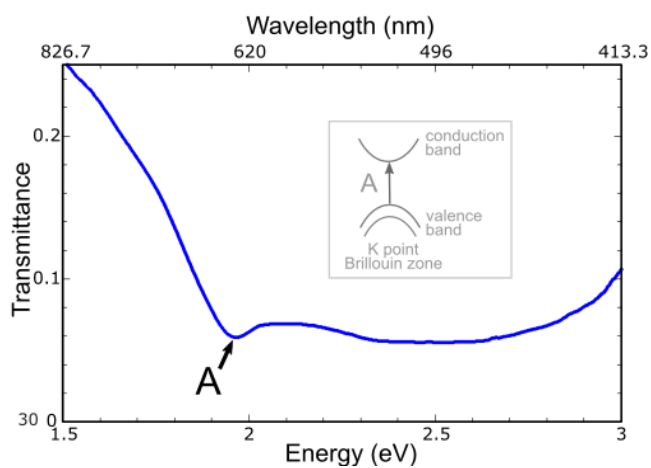

**Figure S3. Transmittance spectrum of a WS₂ film on paper.** The differential transmittance (normalized with respect to bare, uncovered paper) of the WS₂ film shows a prominent peak at ~1.95 eV that agrees well with the one observed in multilayer WS₂ flakes. This peak is attributed to the resonant absorption for photons with the energy matching the direct band gap transition A at the K point of the Brillouin zone.

# Raman characterization of the graphite electrode drawn on top of the films

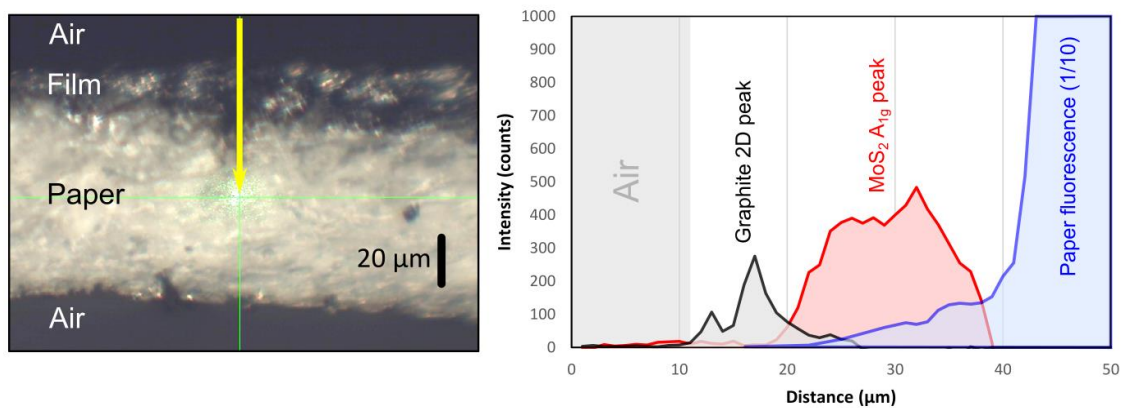

**Figure S4. Characterization of the interface of a graphite electrode drawn onto a MoS₂ film on paper.** (left) Optical microscopy image of the film/paper cross-section studied by Raman spectroscopy. The yellow arrow indicates the approximate line scan measured. (right) Intensity of the different Raman peaks as a function of the distance. The line scan starts outside the sample and at ~10 μm the signal from the graphite 2D peak starts to be measurable. The MoS₂ signal (A$_{1g}$ peak) starts to increase when the graphite signal drops. The paper fluorescence signal is also plotted showing how its intensity increases dramatically once the MoS₂ signal drops. The spot-size used in this measurement is ~5 μm in diameter which would explain the intermixed signal between the graphite and MoS₂.

# Characterization of the electrical properties of a WS$_2$ film on paper

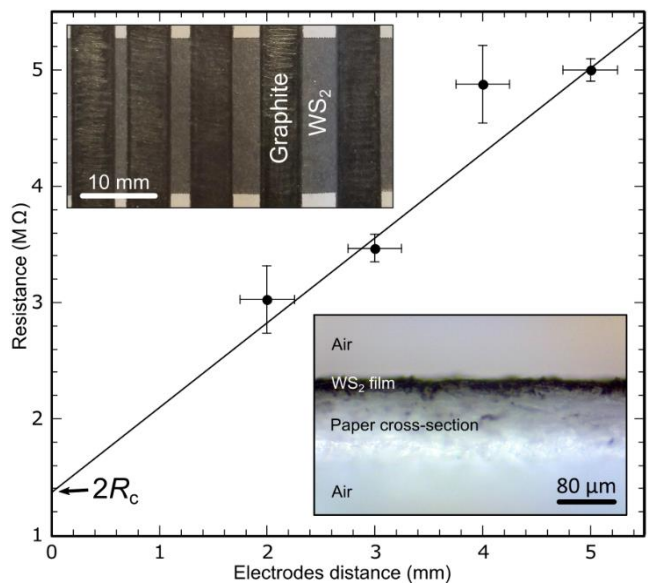

**Figure S5. Transfer length measurement to determine the contact resistance and conductivity of a WS$_2$ film on paper.** A long bar-shaped WS$_2$ film is drawn on paper (20 mm wide, 20 ± 5 μm thick) with graphite electrodes with different electrode spacing. The insets show a picture of the device and a cross-section optical microscopy image of a cross-section of the WS$_2$ film on paper to determine the thickness. The resistance between different pairs of electrodes are measured. The contact resistance ($R_c$ = 0.7 MΩ) is extracted from the crossing of the linear fit with the vertical axis. The conductivity of the film (G = 3.5 ± 1.3 mS/m) is extracted from the slope of the resistance *vs.* electrodes distance linear trend and the device geometry.

# Resistance vs. temperature of device shown in Figure 4 in log scale

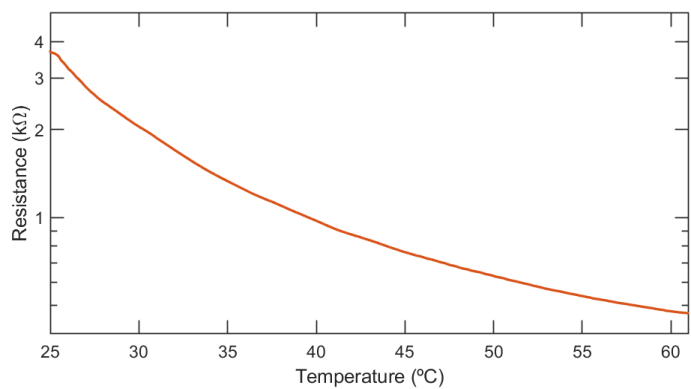

**Figure S6.** Resistance vs. temperature characteristics of the thermoresistive device shown in Figure 4 of the main text but in log-scale to demonstrate that it cannot be fitted to a single exponential decay.